\documentclass[prl,twocolumn,superscriptaddress,floats,showpacs]{revtex4}

\usepackage[dvips]{graphicx}
\usepackage{amsmath}
\usepackage{booktabs}
\usepackage{dcolumn}
\usepackage{sidecap}

\usepackage{rotating}

\newcolumntype{d}{D{.}{.}{-1}}

\newcommand{\tc }{$T_\text{C}$}

\newcommand{\muB}{$\mu_B$}

\begin{document}


\title{Incommensurate magnetic excitations in superconducting LiFeAs}

\author{N.~Qureshi}
\affiliation{II. Physikalisches Institut, Universit\"{a}t zu
K\"{o}ln, Z\"{u}lpicher Str. 77, D-50937 K\"{o}ln, Germany}
\author{P. Steffens}
\affiliation{Institut Laue Langevin, BP156X, 38042 Grenoble Cedex,
France}
\author{Y.~Drees}
\author{A.~C.~Komarek }
\affiliation{II. Physikalisches Institut, Universit\"{a}t zu
K\"{o}ln, Z\"{u}lpicher Str. 77, D-50937 K\"{o}ln, Germany}

\author{D. Lamago}
\affiliation{Laboratoire L\'eon Brillouin, C.E.A./C.N.R.S.,
F-91191 Gif-sur-Yvette Cedex, France} \affiliation{Institut f\"ur
Festk\"orperphysik, Karlsruher Institut f\"ur Technologie (KIT),
Postfach 3640, D-76121 Karlsruhe, Germany}

\author{Y. Sidis}
\affiliation{Laboratoire L\'eon Brillouin, C.E.A./C.N.R.S.,
F-91191 Gif-sur-Yvette Cedex, France}

\author{L.~Harnagea}
\author{H.-J. Grafe}
\author{S.~Wurmehl}
\author{B.~B\"uchner}
\affiliation{Leibniz-Institute for Solid State Research,
IFW-Dresden, 01171 Dresden, Germany}
\author{M.~Braden}
\email{braden@ph2.uni-koeln.de}%
\affiliation{II. Physikalisches Institut, Universit\"{a}t zu
K\"{o}ln, Z\"{u}lpicher Str. 77, D-50937 K\"{o}ln, Germany}

\date{\today}


\begin{abstract}

Magnetic correlations in superconducting LiFeAs were studied by
elastic and by inelastic neutron scattering experiments. There is
no indication for static magnetic ordering but inelastic
correlations appear at the incommensurate wave vector
$(0.5\pm\delta,0.5\mp\delta,0)$ with $\delta \sim$0.07 slightly
shifted from the commensurate ordering observed in other
FeAs-based compounds. The incommensurate magnetic excitations
respond to the opening of the superconducting gap by a transfer of
spectral weight.

\end{abstract}
\maketitle

Superconductivity in the FeAs-based materials \cite{1} appears to
be closely related to magnetism as the superconducting state
emerges out of an antiferromagnetic phase by doping \cite{1,2,3,4}
or by application of pressure \cite{5}. The only FeAs-based
exception to this behavior has been found in LiFeAs, which is an
ambient-pressure superconductor with a high \tc ~ of $\sim$17\ K
without any doping \cite{6,7,8}.
LiFeAs exhibits the same FeAs layers as the other materials but
FeAs$_4$ tetrahedrons are quite distorted \cite{8} suggesting a
different occupation of orbital bands. Indeed ARPES studies on
LiFeAs find an electronic band structure different from that in
LaOFeAs or BaFe$_2$As$_2$ type compounds \cite{9}. The
Fermi-surface nesting, which is proposed to drive the spin-density
wave (SDW) order in the other FeAs parent compounds, is absent in
LiFeAs \cite{9} suggesting that this magnetic instability is less
relevant. The main cause for the suppression of the nesting
consists in the hole pocket around the zone center which is
shallow in LiFeAs \cite{10}. In consequence, there is more density
of states near the Fermi level which might favor a ferromagnetic
instability. Using a three-band model Brydon et al. \cite{10} find
this ferromagnetic instability to dominate and discuss the
implication for the superconducting order parameter proposing
LiFeAs to be a spin-triplet superconductor with odd symmetry.
However, other theoretical analyzes of the electronic
band-structure still find an antiferromagnetic instability which
more closely resembles those observed in the other FeAs-based
materials \cite{10a}.

\begin{figure}
\begin{center}
\rotatebox{270}{
\includegraphics*[width=0.94\columnwidth,angle=90]{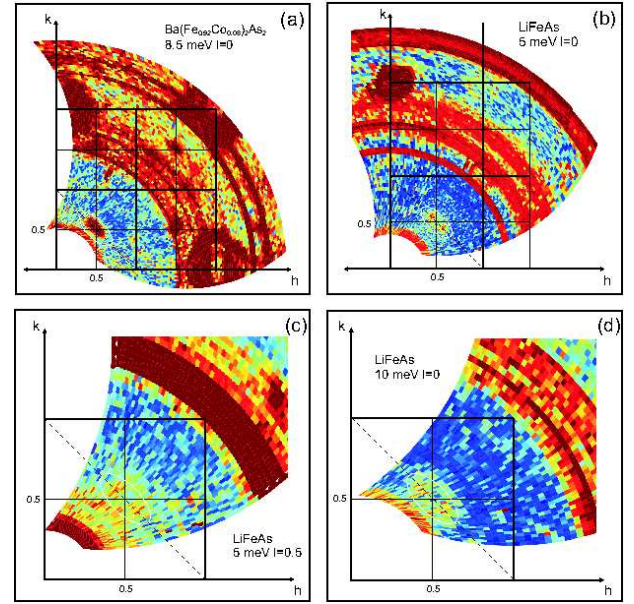}
}
\end{center}
\caption{(color online) Distribution of  neutron-scattering
intensity measured with the flatcone detector on IN20. The energy
transfer is constant in all maps. (hkl) planes with fixed but
finite l-component are studied by tilting the detector and the
sample. Rings of scattering arise from polycristalline
construction material. (a) The scattering  of optimally Co-doped
BaFe$_2$As$_2$ crystal in the superconducting phase (T=10\ K). One
easily identifies the magnetic mode near Q=(0.5,0.5,0) which is of
comparable strength as the phonon scattering around the strong
nuclear Bragg peaks. b-d) Intensity maps measured on LiFeAs at an
energy transfer of 5\ meV and l=0 (T=2\ K), of 5\ meV and l=0.5
(T=2\ K)and of 10 \ meV and l=0 (T=22\ K) for parts b), c) and d),
respectively.
} \label{flatcone}
\end{figure}

Inelastic neutron scattering (INS) experiments revealed magnetic
order and magnetic excitations in many FeAs-based families
\cite{2,11,12,13}. Strong magnetic correlations persist far beyond
the ordered state, and, most importantly, the opening of the
superconducting gap results in a pronounced redistribution of
spectral weight \cite{12,13,14}, which is frequently interpreted
in terms of a resonance mode. Recently a powder INS experiment on
superconducting LiFeAs reported magnetic excitations to be rather
similar to those observed in the previously studied materials
\cite{21} but with a spin gap even in the normal-conducting phase.
Magnetic excitations observed in a recent single-crystal INS study
on non-superconducting Li deficient Li$_{1-x}$FeAs (x$\sim$0.06)
were described by spin-waves associated with commensurate
antiferromagnetism, again with a large temperature independent
spin gap of 13\ meV \cite{wang}. We have performed INS experiments
on superconducting single-crystalline LiFeAs finding
incommensurate magnetic correlations which still can be associated
with the SDW order in the other FeAs-based compounds. These
incommensurate excitations show a clear response to the
superconducting phase.

Single crystals of LiFeAs were grown similarly as described in
reference \onlinecite{15}. Further information documenting the
good chemical, crystalline, and superconducting properties of our
sample crystals is given in the supplementary information. First
neutron experiments were performed with small samples containing
natural Li (about 12$\times$12$\times$0.3mm$^3$) focussing on
elastic analyzes. We searched for magnetic superstructure peaks in
particular near the propagation vectors of the known SDW order in
FeAs-based compounds \cite{2}. With the high sensitivity of
single-crystal neutron diffraction we may rule out this SDW
ordering with a magnetic moment larger than 0.07\muB , which is
significantly below the ordered moment for example in LaOFeAs
\cite{16}. None of the scans along main-symmetry directions
indicates magnetic ordering. For most of the INS experiments we
used two large co-aligned crystals of a total weight of 1.4\ g
grown with the $^7$Li isotope to reduce the neutron absorption.
INS experiments were performed at the thermal spectrometers IN20
(ILL), 1T and 2T (both LLB) as well as on the cold spectrometers
4F (LLB) and IN14 (ILL). On IN20 we used the flatcone detector
with silicon (111) analyzers fixing the final energy to 18.7 meV.

\begin{figure}
\begin{center}
\includegraphics*[width=0.85\columnwidth,angle=0]{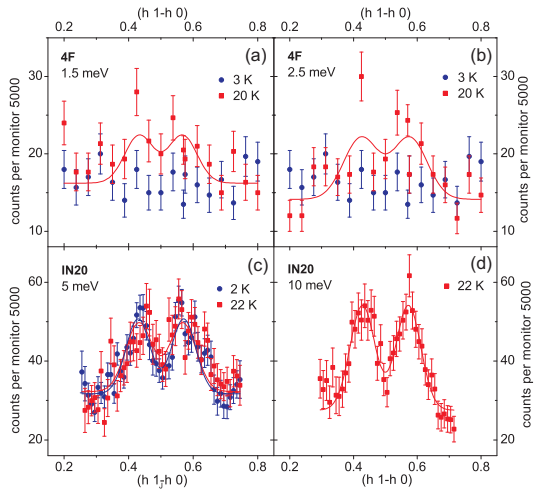}
\end{center}
\vskip -.001 true cm \caption{(Color online) Constant-energy scans
across the incommensurate positions of magnetic scattering
$Q_{inc}=(0.5+\delta,0.5-\delta,0)$ with $\delta \sim \pm$0.07.
The scans in (a) and (b) were recorded on the cold 4F spectrometer
with $k_f$=1.55\AA$^{-1}$ with an energy transfer of 1.5meV and
2.5meV, respectively. One clearly recognizes the incommensurate
signal even at this low energy in the normal state, but the signal
disappears in the superconducting phase. Parts (c) and (d) show
scans taken from the flatcone data at 5\ meV and 10\ meV,
respectively. Lines are fits with two symmetric Gaussians. }
\label{const-e-scans}
\end{figure}

Fig. 1 shows the maps recorded with the flatcone detector on IN20.
Fig. 1 (a) represents the results for Co-doped BaFe$_2$As$_2$ in
the superconducting phase. This pattern demonstrates, how easily
this instrument may detect the magnetic signal which is found to
be similarly strong to that of the phonon scattering around the
nuclear Bragg peaks. The same experiment on LiFeAs shown in Fig.
1(b) immediately reveals the differences of the magnetic
scattering. Although the phonons in both samples yield signals of
similar strength, there is no comparably strong magnetic signal
visible in LiFeAs. However, there clearly is magnetic scattering
near $Q$=(0.5,0.5,0) which is displaced in the transverse
direction to $Q_{inc}=(0.5\pm\delta,0.5\mp\delta,0)$ with $\delta
\sim$0.07. Throughout the paper we label all reciprocal-space
vectors in reduced lattice units referring to the lattice with a
3.8 \AA ~ parameter.
The incommensurate excitation is also visible in the pattern
obtained with l=0.5 at 5\ meV energy transfer and in the l=0
pattern with 10\ meV energy transfer, see Fig. 1 (c) and (d). The
observation of a comparably strong signal for finite l suggests an
essentially two-dimensional magnetic correlation, therefore we
neglect a possible out-of-plane modulation in the following
discussion. By normalizing with the phonon signals, we may compare
the magnetic scattering in LiFeAs and in Co-doped BaFe$_2$As$_2$ .
Taking into account the scattering lengths and the
reciprocal-energy factor and assessing the phonon dispersion of
frequency and of dynamical structure factors with the aid of
phenomenological lattice-dynamics models, we may roughly estimate
the incommensurate magnetic signal per Fe in LiFeAs at 5\ meV and
2\ K to be a factor eight less intense than the commensurate
scattering in Co-doped BaFe$_2$As$_2$ appearing at 8.5\ meV and
10\ K \cite{14}, but note that integration over the Brillouin zone
will recover a factor two. The constant-energy maps do not give
any indication of a ferromagnetic scattering in LiFeAs.

Scans across the incommensurate magnetic signal were obtained from
the IN20 scattering maps and by additional experiments on cold and
thermal triple-axis spectrometers. A few examples of  transverse
scans, (h 1-h 0), are shown in Fig. 2. The profiles were fitted
with two symmetric Gaussians appearing at
$Q_{inc}=(0.5\pm\delta,0.5\mp\delta,0)$ with $\delta\sim$0.07, to
extract the incommensurability and the amplitude of the magnetic
signal. In the normal-conducting state we were able to follow the
incommensurate signal between 1.5 and 10 meV finding almost no
energy dependence of the incommensurability $\delta$, see Fig.
3(a). Combining the data obtained on the thermal and on the cold
triple-axis spectrometers we also obtain the energy dependence of
the strength of this signal. After correcting for the monitor and
for the Bose factor we may deduce the imaginary part of the
generalized susceptibility, $\chi ''(Q_{inc},E)$, which is shown
in Fig. 3(b) for the temperature of 22\ K. $\chi ''(Q_{inc},E)$
can be well described by a single relaxor function relating $\chi
''$ with the real part of the susceptibility at zero energy and a
characteristic energy $\Gamma$: $ \chi ''(Q_{inc},E)=\chi
'(Q_{inc},0) {\Gamma E \over \Gamma^2+E^2}$
 yielding  $\Gamma$=6.0$\pm$0.6\ meV. This rather
low value of the characteristic energy signals that LiFeAs is
quite close to the corresponding SDW instability. The
incommensurate scattering in LiFeAs and its spectrum closely
resemble the incommensurate magnetic excitations arising from
nesting in Sr$_2$RuO$_4$ where the corresponding SDW instability
can be induced by a small substitution \cite{srruo}.

\begin{figure}
\begin{center}
\includegraphics*[width=0.85\columnwidth,angle=0]{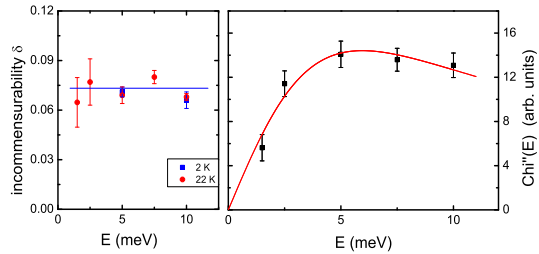}
\end{center}
\vskip -.001 true cm \caption{(Color online) (a) Energy dependence
of the incommensurability of magnetic scattering in LiFeAs. (b)
The energy dependence of the amplitude of the signal can be well
described by a single relaxor function (line), see text, with a
characteristic energy of $\Gamma$=6.0$\pm$0.6\ meV signalling a
near SDW instability. } \label{fig3}
\end{figure}

The comparison of the constant-energy scans above and below the
superconducting transition reveals a pronounced shift of spectral
weight associated with the superconducting transition in LiFeAs.
The incommensurate signals at 1.5 and 2.5\ meV become almost fully
suppressed in the superconducting phase. In contrast there is
evidence for an increase in intensity at higher energies. In order
to elucidate this transfer of spectral weight we performed
constant-Q scans at the position of the incommensurate signal,
$Q_{inc}$, which are shown in Fig. 4. At low energy in the
superconducting phase the scattering at $Q_{inc}$ is suppressed to
the background, while the signal is enhanced in the energy range 6
to 10 meV. With the present statistics there is no indication for
magnetic scattering persisting in the superconducting state for
E$<$4\ meV suggesting a clean gap in the magnetic excitations in
the superconducting state. ARPES and specific heat measurements on
LiFeAs indicate two gaps opening in the bands in LiFeAs which
amount to $2\Delta_1$=2.4 meV and $2\Delta_2$=5.2 meV \cite{18}.
The higher gap value agrees with our observation that transfer of
spectral weight from below 4.5\ meV to above 4.5\ meV occurrs upon
entering the superconducting phase. Due to the limited statistics
of the temperature-dependent data shown in Fig. 5 we may not yet
fully determine the relation between the spectral-weight shift and
the superconducting transition. But the data, in particular that
in Fig. 5(b), unambiguously confirm that the transfer of spectral
weight represents the response of the system to the opening of the
superconducting gaps. By measuring the depolarization of the
polarized neutron beam due to the shielding of the guide fields we
may ascertain the good homogeneity of the superconducting phase in
the large superconducting single crystals (T$_c$=16.4\ K).

The experimental data obtained by powder INS and the given
interpretation \cite{21} only qualitatively agree with our
experiment. The incommensurate character of magnetic excitations
in LiFeAs is not easily accessible in a powder experiment, but our
results fully disagree with the claim of a spin gap in the normal
state that is formulated in reference \onlinecite{21}. We
speculate that the background and phonon contributions underlying
the magnetic signal could not be properly assessed thereby
underestimating the magnetic response of LiFeAs.

Magnetic excitations in LiFeAs clearly differ from those reported
for many other FeAs-based superconductors in at least two aspects.
The amplitude of the signal is weaker than that in Co-doped
BaFe$_2$As$_2$. More importantly the signal clearly appears at an
incommensurate position, whereas those in the FeAs superconductors
with a high \tc \ are all commensurate. Incommensurate scattering
has recently been observed in the end member of the
Ba$_{1-x}$K$_x$Fe$_2$As$_2$ series KFe$_2$As$_2$ \cite{chul-ho}
which, however, exhibits a very low \tc .

The incommensurability in the magnetic scattering seems to be the
consequence of the suppressed nesting condition in LiFeAs. In the
plots relating the superconducting transition temperature with
either the As-Fe-As bond angles \cite{19} or with the anion height
from the Fe layer \cite{20}, LiFeAs clearly lies on the wing of
the distribution away from the regular tetrahedron observed in
samples with maximum \tc . The sizeable tetrahedron elongation in
LiFeAs should cause a different orbital occupation. It appears
interesting to note, that concerning these distortions LiFeAs
resembles FeTe$_{\sim0.5}$Se$_{\sim0.5}$ which exhibits similar
incommensurate excitations \cite{fetese}. It appears likely that
doping and structural deformation change the occupation of orbital
levels and thereby the character of the magnetic instability.
DFT calculations indicate that doping by electrons or by holes
modifies the nesting with the magnetic response shifting from the
commensurate position to an incommensurate one
\cite{park,yaresko}. Hole doping implies a longitudinal modulation
peaking at $Q=(0.5\pm\delta,0.5\pm\delta,0)$ which indeed is
observed in hole-overdoped KFe$_2$As$_2$ \cite{chul-ho}, whereas
electron doping results in a transverse modulation peaking at
$Q=(0.5\pm\delta,0.5\mp\delta,0)$ \cite{park,yaresko}.
Experimental evidence for such transverse inelastic
incommensurability can so far only be found in the high-energy
magnetic excitations in Co-doped BaFe$_2$As$_2$ \cite{ppppp},
which however are associated with commensurate scattering at lower
energy. After submission of our article, a static transverse
modulation was reported for under-doped
Ba(Fe$_{1-x}$Fe$_x$)$_2$As$_2$ \cite{29}. Our observation of
transversally modulated incommensurate excitations in LiFeAs thus
suggests to compare LiFeAs with an electron doped compound. The
similarity can arise from the role of the central hole pocket
which is shallow in LiFeAs somehow similar to the expected effect
of electron doping. The transversally modulated incommensurate
response in LiFeAs, however, seems still to be closely related
with the commensurate or longitudinally modulated response in the
other FeAs-based materials.

\begin{figure}
\begin{center}
\includegraphics*[width=0.85\columnwidth,angle=0]{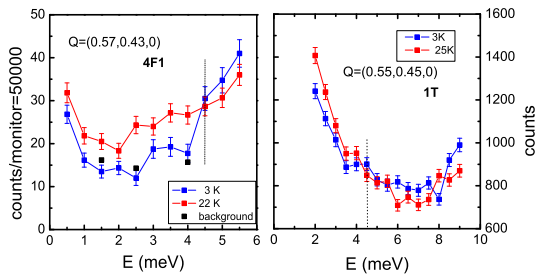}
\end{center}
\vskip -.001 true cm \caption{(Color online) Energy dependence of
the INS intensity at $Q_{inc}$ at temperatures above and below the
superconducting transition. Part(a) shows the raw intensity
measured on the cold spectrometer (with high energy resolution,
$k_f$=1.55\AA$^{-1}$), while part (b) shows data measured on the
thermal spectrometer with lower resolution. Vertical bars indicate
the energy of the crossover of the shift of spectral weight.}
\label{escans}
\end{figure}

\begin{figure}[!t]
\begin{center}
\includegraphics*[width=0.95\columnwidth,angle=0]{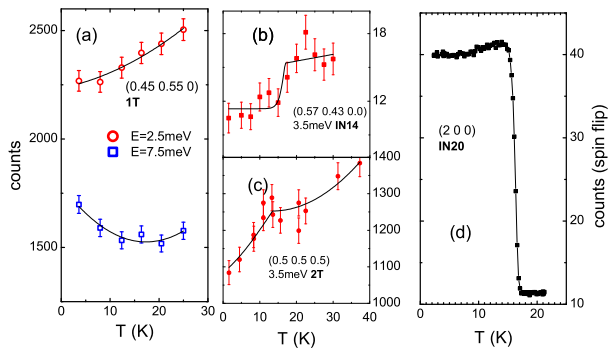}
\end{center}
\vskip -.001 true cm \caption{(Color online) Temperature
dependence of scattering intensity at $Q_{inc}$ for fixed energy
transfers, data are measured on the 1T spectrometer (a) and on
IN14 (b) with the two large crystals ; data in (c) were taken with
a smaller crystal on 2T. Lines are guides to the eye. (d)
Temperature dependence of spin-flip scattering at the nuclear
Bragg peak (200) reflecting the neutron depolarization due to
superconducting shielding.} \label{fig5}
\end{figure}

Inspection of the Fermi surfaces calculated in reference
\onlinecite{10a} or those fitted to the ARPES data \cite{10}
allows one to understand that the commensurate wave vector
(0.5,0.5,0) is not associated with the strongest magnetic signal
in LiFeAs as such nesting is absent in this material. However, an
additional shift may partially recover nesting. Taking the orbital
character of the Fermi-surface sheets into account the
experimentally determined transversal incommensurability of
$\delta \sim$0.07 agrees with the Fermi-surface maps presented in
references \onlinecite{10,10a}, but a detailed calculation is
desirable.

In conclusion INS experiments on single-crystalline
superconducting LiFeAs reveal incommensurate magnetic
correlations, which appear close to the wave vector of the
stronger magnetic signal observed in previously studied FeAs-based
superconductors. The loss of commensurate nesting for
q=(0.5,0.5,0) apparently needs to be compensated by a small shift
explaining the incommensurate propagation vector
$Q_{inc}=(0.5\pm\delta,0.5\mp\delta,0)$ with $\delta\sim$0.07.
These magnetic fluctuations clearly respond to the opening of the
superconducting gap. In the superconducting phase the magnetic
weight at $Q_{inc}$ seems to be fully suppressed below $\sim$5\
meV and there is an enhancement of spectral weight compared to the
normal state in the energy range 6 to 10\ meV. The magnetic
instability in LiFeAs indicates that magnetic correlations in FeAs
based superconductors are more variable than a simple commensurate
response.

\par This work was supported by the Deutsche Forschungsgemeinschaft
through SFB 608 and through the Priority Programme SPP1458 (Grant
No. BE1749/13). We thank C. H. Lee, I.~Morozov, S.~Aswartham, and
C.~Nacke for various discussions, A.U.B. Wolter for a
magnetization measurement, L. Giebeler for the use of x-ray
equipment, and C.-H. Lee for providing a large single crystal of
Co-doped BaFe$_2$As$_2$.

\end{document}